\documentclass{cernyrep}

\usepackage{varwidth}
\usepackage{xcolor}

\begin{document}
\title{Laser Wakefield Accelerators}
\author{Z. Najmudin}
\institute{John Adams Institute for Accelerator Science, The Blackett Laboratory, Imperial College London, UK}
\maketitle

\begin{abstract}
The one-dimensional wakefield generation equations are solved for increasing levels of non-linearity, to demonstrate how they contribute to the overall behaviour of a non-linear wakefield in a plasma. The effect of laser guiding is also studied as a way to increase the interaction length of a laser wakefield accelerator.\\

\end{abstract}

\section{Introduction}

Large amplitude relativistic plasma waves, driven in the wake of a relativistic driver passing through a plasma, are a potential linear accelerator system \cite{tajima}. This is commonly called a wakefield accelerator. The plasma, being already ionized, can support accelerating fields many orders of magnitude greater than in conventional accelerators. Indeed, at the time of writing, reports of energy gains of ${\sim}1\UGeV$ in distances of centimetres are now becoming common \cite{kneip, texas, taiwan, leemans}.

Possible drivers of the relativistic plasma wave include light \cite{tajima}, electron \cite{pchen}, positron \cite{blue}, or proton beams \cite{caldwell}. A generalized one-dimensional (1D) wakefield generation equation describing generation by all of the mentioned drivers is \cite{esarey}
\begin{equation}
\frac{1}{{k_\mathrm{p}}^{2}}  \frac{\partial^{2} \phi }{\partial \zeta^{2}}  = \pm \frac{n_\mathrm{b}}{n_{0}} + {\gamma_\mathrm{p}}^{2} \left\{ \beta_\mathrm{p}\left[1- \frac{(1+a^{2})}{{\gamma_\mathrm{p}}^{2}(1+\phi)^{2}} \right]^{-1/2} - 1 \right\},
\end{equation}
where the potential $\phi$ is normalized to $mc^{2}/e$ and the variations are in the quasi-static frame with $\zeta~=~z - v_\mathrm{p}t$ for a driver of velocity $v_\mathrm{p}$. The first term on the right describes the response to a particle beam driver and is $+$ for an electron and $-$ for a positively charged driver. In its absence ($n_\mathrm{b}=0$), and in the limit that the phase velocity of the wave $\gamma_\mathrm{p}\gg1$, then expanding the square brackets and also expanding $\beta_\mathrm{p} = (1-\frac{1}{{\gamma_\mathrm{p}}^{2}})^{1/2}$ gives a non-linear laser wakefield driver equation:
\begin{equation}
\frac{1}{{k_\mathrm{p}}^{2}}  \frac{\partial^{2} \phi }{\partial \zeta^{2}}  =  \frac{1}{2} \left[ \frac{(1+a^{2})}{(1+\phi)^{2}} -1 \right]
\label{eqn:linear}.
\end{equation}
In the small amplitude limit, $\phi \ll 1$, this can be further simplified to
\begin{equation}
\left( \frac{\partial^{2}  }{\partial \zeta^{2}}  + k_\mathrm{p}{^{2}}\right) \phi= \frac{1}{2}{k_\mathrm{p}}^{2}a^{2}
\label{eqn:driven},
\end{equation}
which reveals itself to be a driven oscillator. Below, we obtain \Eref{eqn:driven} starting from the linear forms of the fluid equations of motion, continuity and Gauss's Law. The electric field and density can then be obtained from ${\bf E} = -\nabla \phi$ and Poisson's equation. This allows us to obtain some of the common characteristics of a laser driven wakefield accelerator. By adding further levels of non-linearity, we reveal some of the behaviour of laser wakefield accelerators in the non-linear regime.

\section{Laser wakefield}

\subsection{Basic equations}

Beginning with Gauss's Law, and the continuity and fluid motion equations for electrons in a plasma in one dimension (the ions are assumed to be heavy and thus fixed) subject to a laser of peak intensity $I$ and wavelength $\lambda$ with a normalized vector potential $a = {eE}/{m\omega c}\approx 8.9\times10^{-6} \left(I\lambda^{2})\right)^{0.5}$ [SI units]:
\[
\epsilon_{0}\frac{\partial E}{\partial z} = - e (n_\mathrm{e} - n_\mathrm{i}), \quad
\frac{\partial n_\mathrm{e}}{\partial t} +  \frac{\partial}{\partial z} {(n_\mathrm{e}  v)} = 0, \quad
\left[ \frac{\partial p}{\partial t} + v \frac{\partial p}{\partial z} \right]= -eE + \frac{1}{\gamma_{\perp}}mc^{2}\frac{\partial (a^2 )}{\partial z},
\]
where $p=\gamma m v$ is the momentum of the fluid element. Note that the last term on the right of the equation of motion is the relativistic ponderomotive force, with the relativistic factor due to the quiver motion usually given by $\gamma_{\perp} = \sqrt{1+(a^{2}/2)}$ for linear polarization. It is this ponderomotive force which mediates the interaction between laser and plasma.

For simplicity, we use normalized units: $\epsilon_{0}, e, m, c =1$. We also move to the frame in which the laser driver is stationary (the quasistatic approximation), i.e. $\zeta = z - v_\mathrm{p}t \approx z - ct$. So $\frac{\partial}{\partial z} = \frac{\partial}{\partial \zeta}$ and $\frac{\partial}{\partial t} = -c\frac{\partial}{\partial \zeta} $. Also using $n_\mathrm{e} = n_{0} + n_{1}$, where $n_{0}$ is the initial plasma density and $n_{1}$ is therefore the amplitude of the plasma wave, the above equations become
\begin{equation}
\frac{\partial E}{\partial \zeta} = - n_{1}, \quad
 n_{1} = (n_{0}+n_{1})  \beta, \quad
(1-\beta)\frac{\partial p}{\partial \zeta} = E - \frac{1}{\gamma_{\perp}}\frac{\partial (a^2 )}{\partial \zeta}.
\label{eqn:nonlinear}
\end{equation}
Here $\beta = v/c$ and is given by
$$\beta = \frac{p}{(1+p^2+a^{2})^{1/2}},$$
i.e. the relativistic factor of the electrons depends on not only their longitudinal motions but also their transverse oscillations, which at the intensities used to drive laser wakefield accelerators are also relativistic.

Starting from these base equations, we can now consider different levels of non-linearity and assess their effect on wakefield generation.

\subsection{Linear wakes}

\begin{figure}[b]
\quad (a) \hspace{4.5cm} (b) \hspace{4.55cm} (c) \vspace{-3mm}
\begin{center}
\includegraphics[width=5cm]{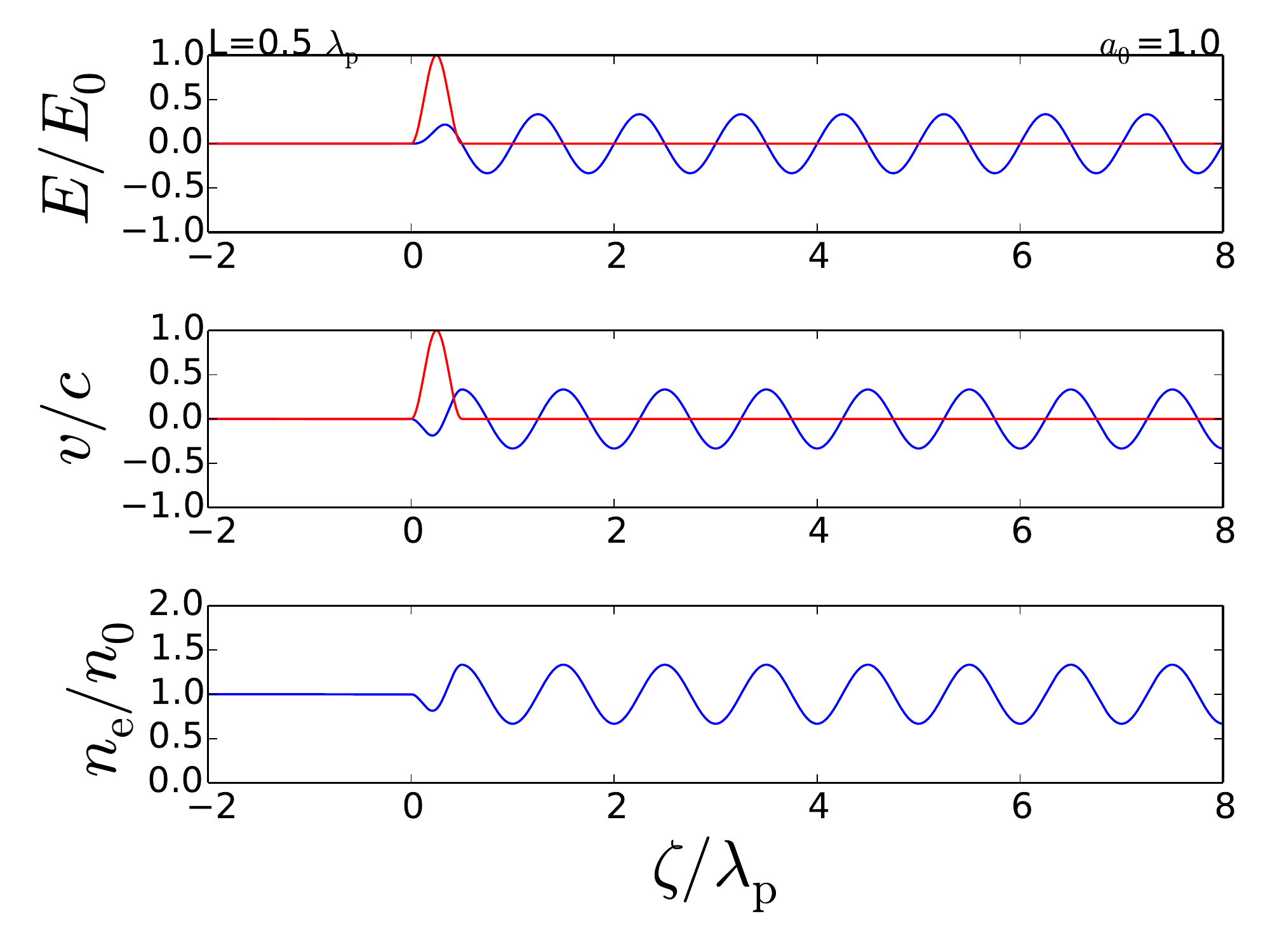}
\includegraphics[width=5cm]{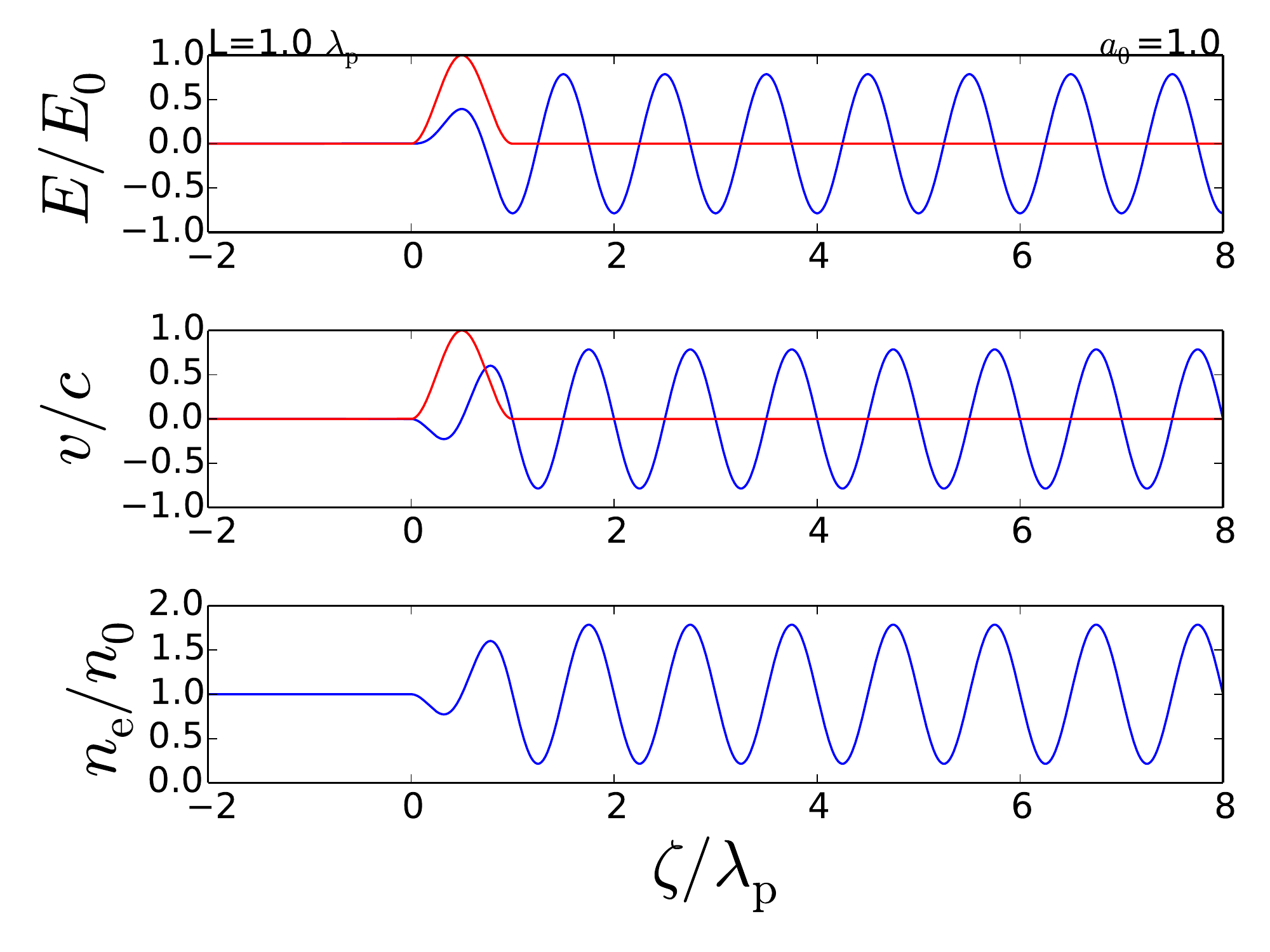}
\includegraphics[width=5cm]{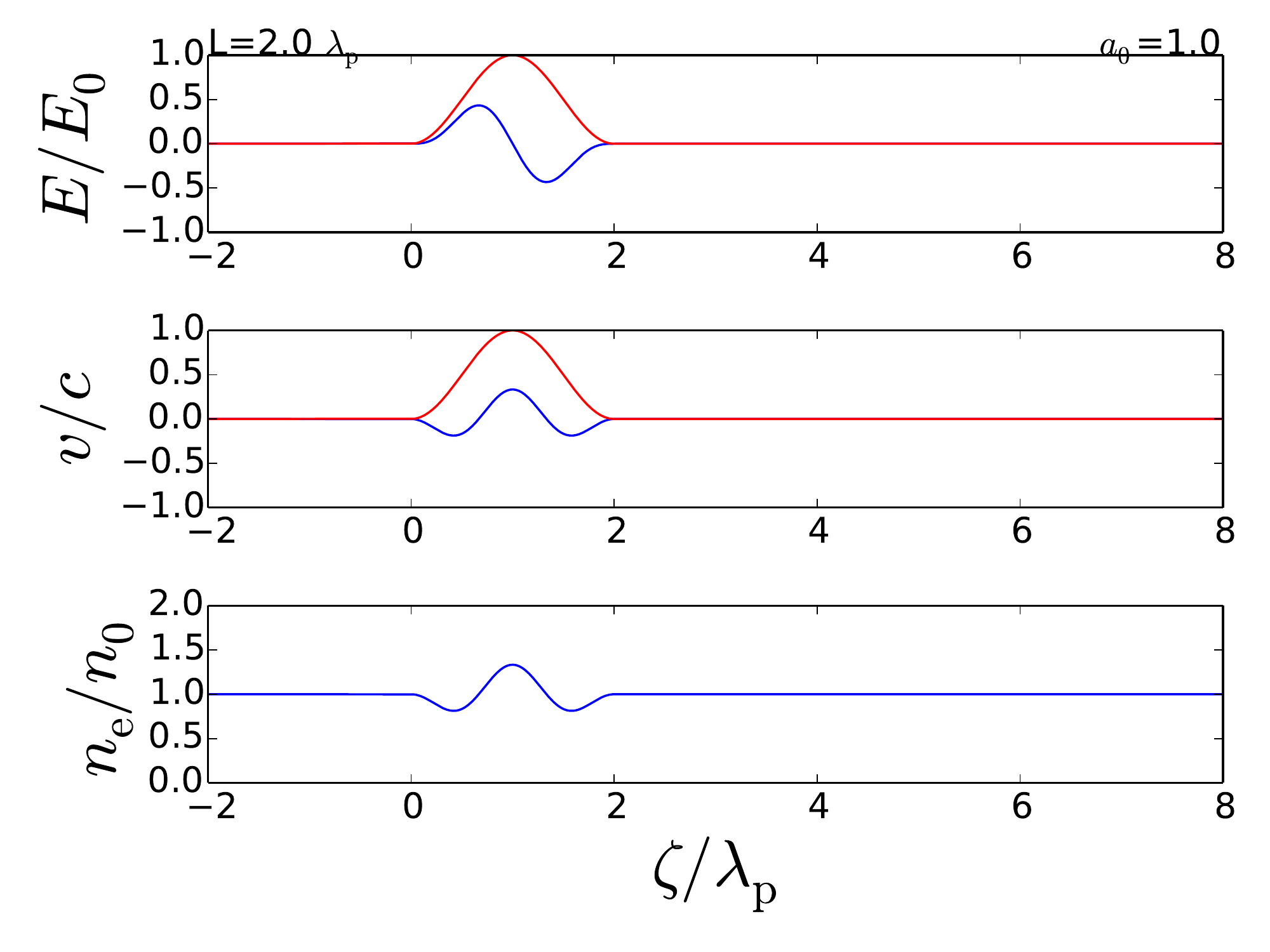}
\caption{Linear wakefield generation: plots of $E/E_{0}$, $v/c$, and $n_\mathrm{e}/n_{0}$ (in blue) for laser pulse $a=a_{0}\sin(\pi \zeta/L)$ with pulse length $L= {\rm (a)}\, \frac{1}{2}, {\rm (b)}\, 1, {\rm (c)}\, 2 \, \lambda_\mathrm{p}$ and $a_{0} = 0.2$ (in red).}
\label{denScan}
\end{center}
\end{figure}

For small amplitudes, we can assume, $n_{1} \ll n_{0}$, $\beta\ll 1$, $\gamma = 1$ and $p = mc \beta$:
\[
\frac{\partial E}{\partial \zeta} = - n_{1}, \quad
 n_{1} = n_{0}  \beta, \quad
\frac{\partial \beta}{\partial \zeta} = E - \frac{\partial (a^2 )}{\partial \zeta}.
\]
Combining these equations gives us essentially \Eref{eqn:linear}, with (in normalized units) $k_\mathrm{p}{^{2}} = n_{0}$. For a normalized laser intensity profile $a^{2} = a_{0}{^{2}}\sin^{2}(\pi \zeta/L)$ for $0 < \zeta < L$, i.e.~a half sinusoid, the wake growth is optimized for $L = \lambda_\mathrm{p} \equiv 2\pi / k_\mathrm{p}$, or alternatively, when the intensity full-width half-max $L_\mathrm{fwhm} = \lambda_\mathrm{p}/2$. This can be seen in \Fref{denScan} where these equations have been solved numerically using the scipy.integrate.odeint function in python. One can see that the plasma wave grows at the front end of the laser pulse as the ponderomotive force pushes electrons ahead of it, which then relax to form a plasma oscillation. When resonant, the back of the pulse gives an extra ponderomotive kick to the plasma wave growing it to higher amplitude.
At resonance, the wake trailing (behind) the laser pulse has solution
\begin{equation*}
n_\mathrm{e} = n_{0} \left(1 +  \frac{\pi}{4} a_{0}{^{2}} \sin k_\mathrm{p} \zeta \right), \qquad
E_{z} = E_{0}  \frac{\pi}{4} a_{0}{^{2}} \cos k_\mathrm{p} \zeta,  \qquad
	\phi = - \frac{mc^{2}}{e}  \frac{\pi}{4} a_{0}{^{2}} \sin k_\mathrm{p} \zeta,
\end{equation*}
where $E_{0} = (mc\omega_\mathrm{p}/e)$ is the electric field for sinusoidal oscillations of maximum amplitude, i.e.~for $ n_{1} = n_{0}$.
For yet longer pulse lengths, the wakefield (the plasma wave amplitude following the laser pulse) reduces again, to the point where for  $L = 2\lambda_\mathrm{p}$, there is no wakefield at all. There is however a driven (or forced) oscillation of the plasma wave whilst the laser pulse is still active (see \Fref{denScan}(c)), which may be useful if only a single bunch needs to be accelerated, rather than a pulse train. There are also smaller amplitude resonances for $L \approx (n+\frac{1}{2})\lambda_\mathrm{p}$ for integer $n>2$.

Clearly there is a problem for linear solutions with $a_{0}>1$, where $n_\mathrm{e}$ can become negative. In fact, in the fluid model, electron sheets would cross, so that instead of seeing a returning force, some electrons end up being continually accelerated, a process called `wavebreaking' \cite{dawson}. Wavebreaking is interesting because it suggests that plasma electrons can themselves be a source of particles to be accelerated in the accelerator \cite{modena}. Thus, this cold linear wavebreaking limit for the electric field $E_{0} = (mc\omega_\mathrm{p}/e)$ is useful for giving an estimate of the energy output from the accelerator. In the laser driven case, the wakefield travels at the group velocity of the driving pulse, as quantified by $\gamma_\mathrm{ph} \approx (\omega_{0}/\omega_\mathrm{p})$ for a laser of angular frequency $\omega_{0}$. The maximum energy gain is then given by
\[
W_\mathrm{max} \approx 2 \gamma_\mathrm{ph}{^{2}} mc^{2},
\]
and so is dependent on the plasma density. Interestingly, lower density (smaller $\omega_\mathrm{p}$, thus higher $\gamma_\mathrm{ph}$) results in greater energy gain. This, however, comes at the expense of the greater laser energy that is required to drive the larger sized wakefield that results from lower density.

\subsection{Non-linear wakes}
\begin{figure}[h]
\quad (a) \hspace{4.5cm} (b) \hspace{4.55cm} (c) \vspace{-3mm}
\begin{center}
\includegraphics[width=5cm]{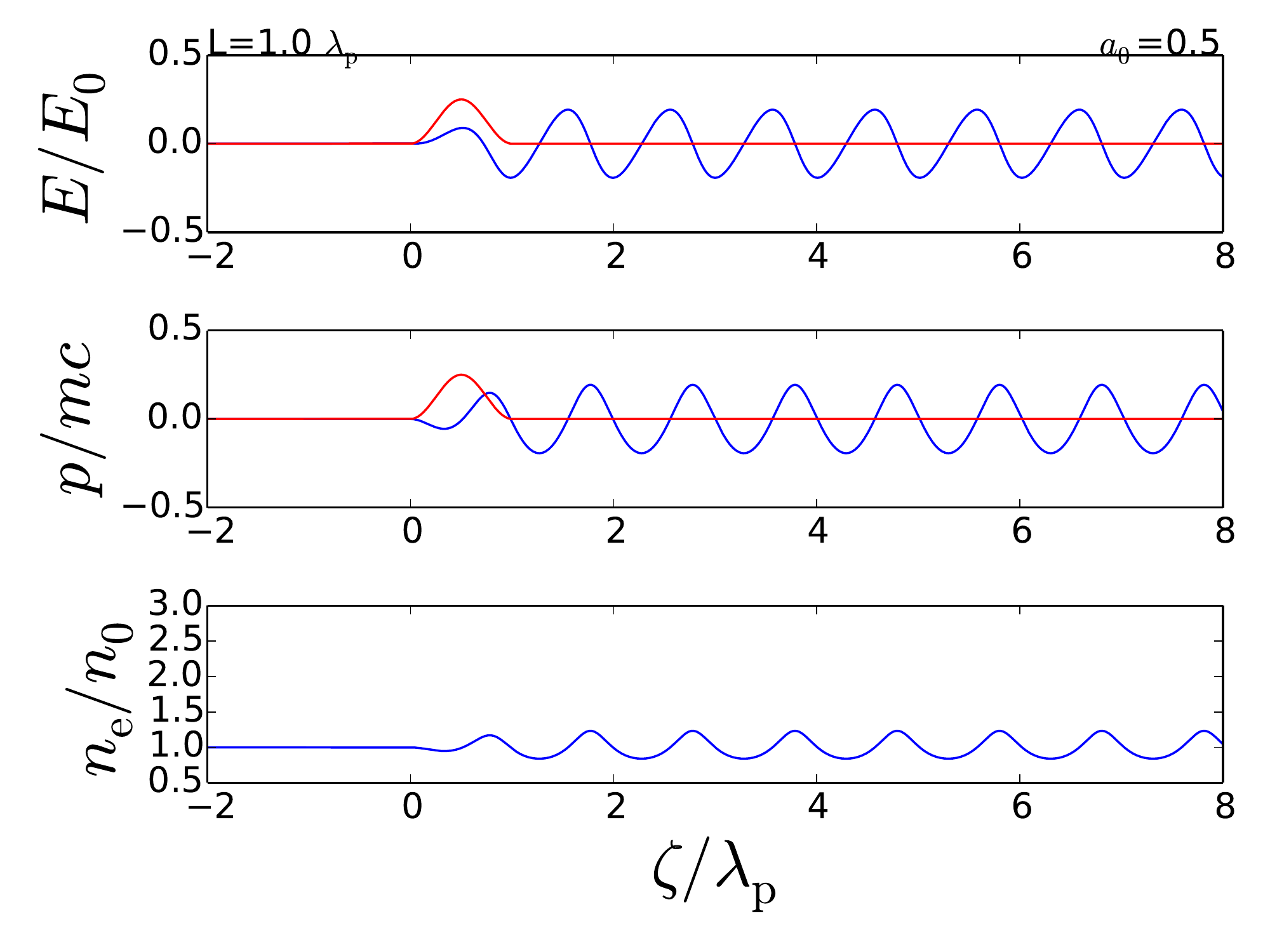}
\includegraphics[width=5cm]{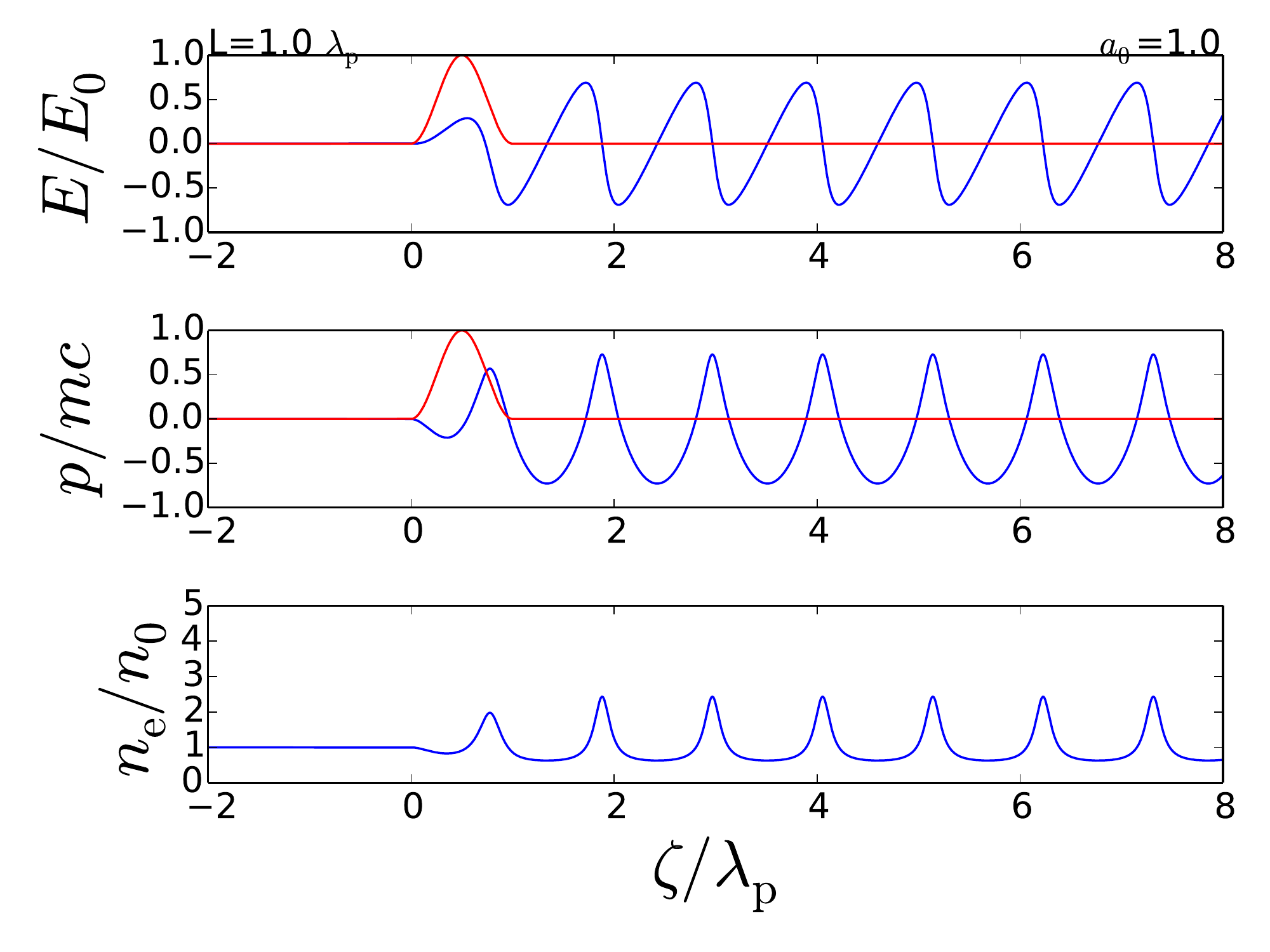}
\includegraphics[width=5cm]{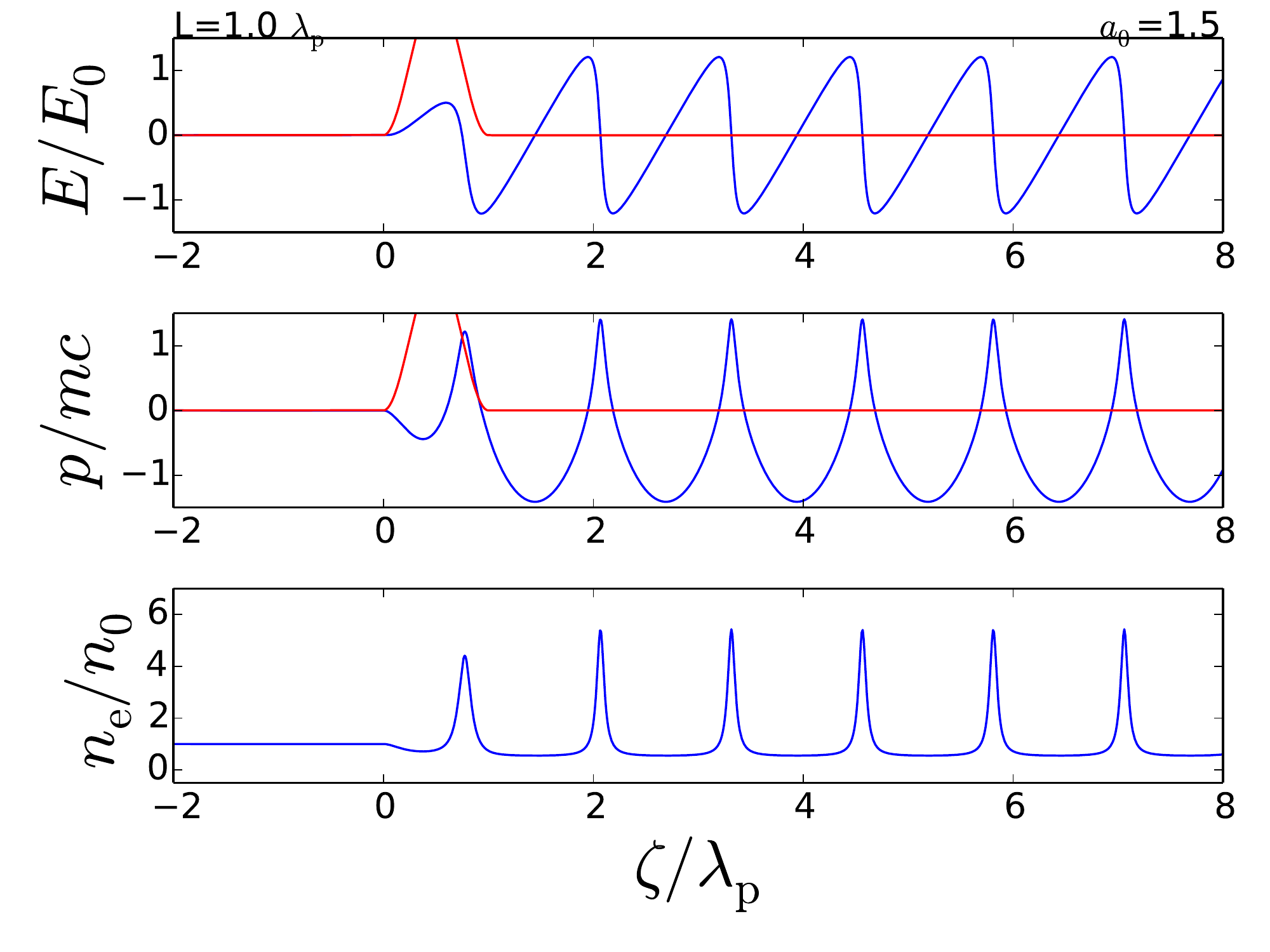}
\caption{Non-linear wakefield generation: plots of $E/E_{0}$, $v/c$, and $n_\mathrm{e}/n_{0}$ (in blue) for laser pulse $a=a_{0}\sin(\pi \zeta/L)$ with $a_{0}= {\rm (a)}\, 0.5, {\rm (b)}\, 1.0$ and ${\rm (c)}\, 1.5$ (in red).}
\label{aScan}
\end{center}
\end{figure}

For $n_{1} \sim n_{0}$, the continuity equation gives $n_{1} = n_{0} (1 - \beta)^{-1}$. Since in their fluid motion, the electrons have $-1 < \beta < 1$, one can see that $\frac{1}{2} < n_{1}/n_{0} < \infty$. Because of this non-linearity in the continuity equation, the lowest density in a 1D wake cannot be less than $\frac{1}{2}n_{0}$. The plasma wave as a result becomes non-linearly steepened, with sharp high-density spikes associated with longer areas of lower electron density plasma.
Of course this is only true in one dimension. In three dimensions, at high $a_{0}$, transverse motion can lead to $n_\mathrm{e} \rightarrow 0$, or cavitation. Relativistic effects in the equation of motion add further non-linearity, as in \Eref{eqn:nonlinear}. The effect of these non-linearities can be seen in \Fref{aScan}, for a half sinusoidal laser pulse (as before) with $L=\lambda_\mathrm{p}$, and increasing laser strength $a_{0}$.

\begin{figure}[h]
\hspace{2.8cm} (a) \hspace{4.5cm} (b) \vspace{-3mm}
\begin{center}
\includegraphics[height = 4.0cm]{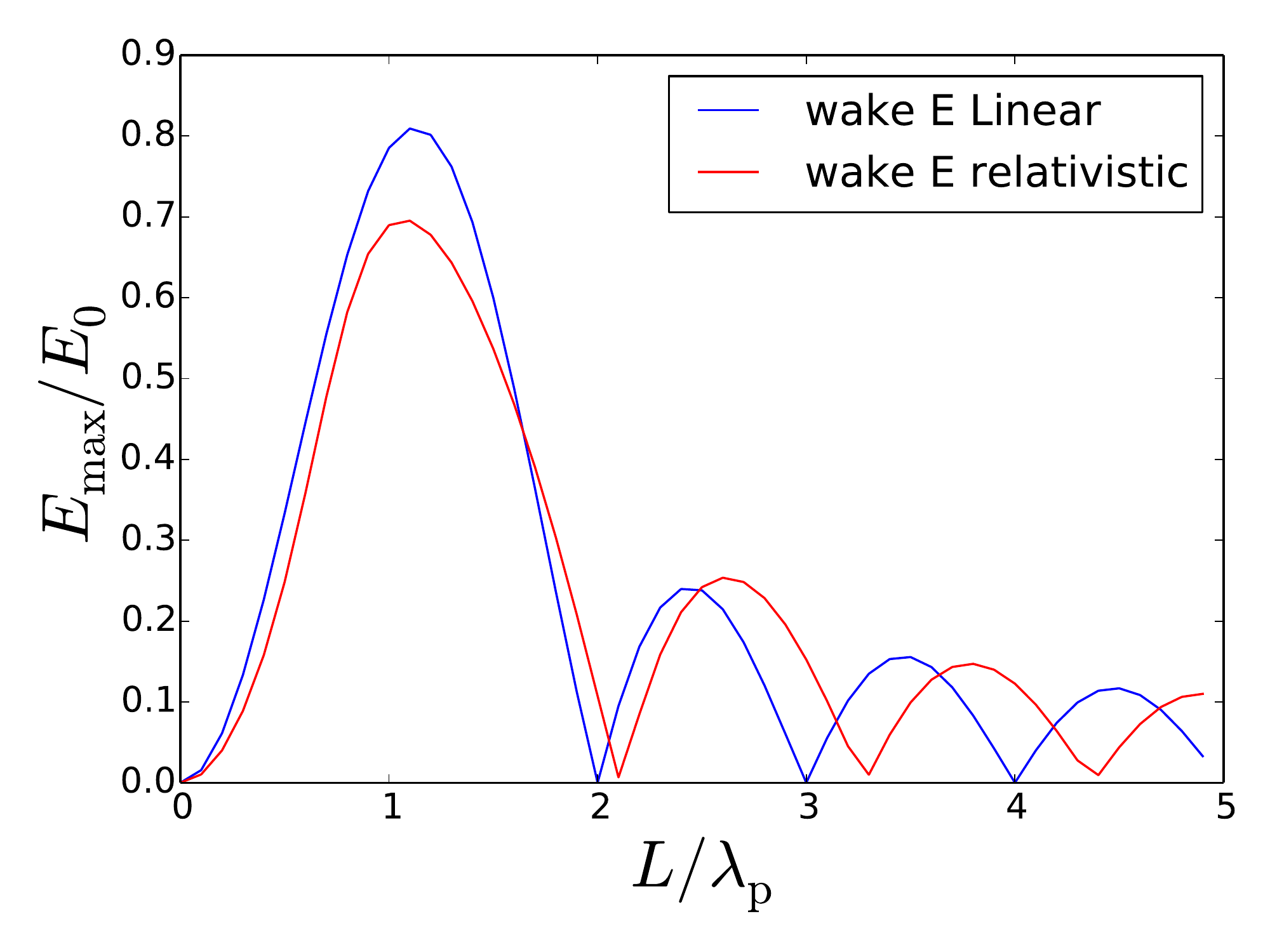}
\includegraphics[height = 4.0cm]{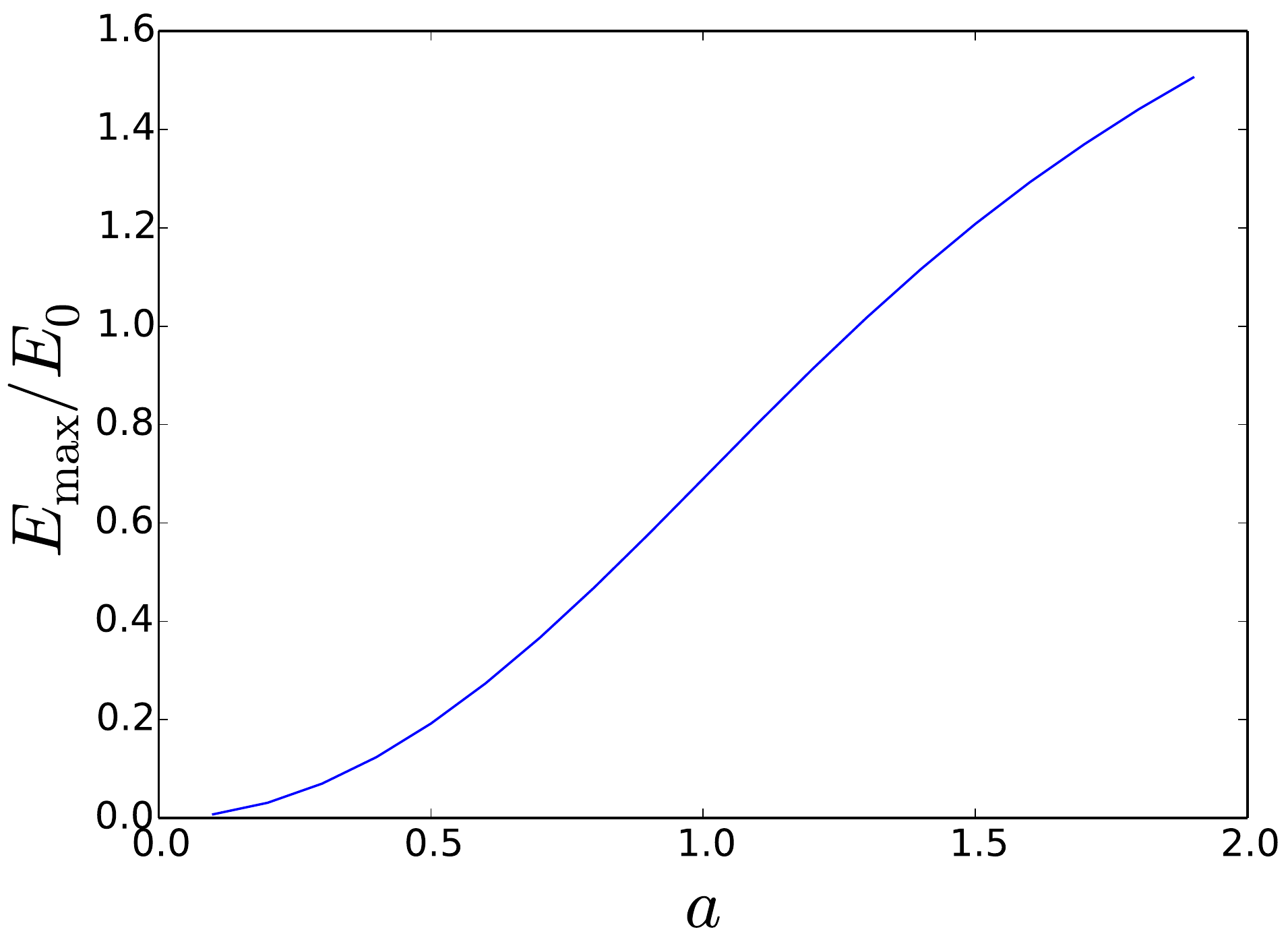}
\caption{Wake amplitude for (a) $a_{0}=1$ with varying pulse length for a linear (blue) and non-linear (red) wake, and (b) $L = \lambda_\mathrm{p}$ and varying $a_{0}$.}
\label{aScan2}
\end{center}
\end{figure}

Even for $a_{0}=0.5$ (\Fref{aScan}(a)), the density profiles of the plasma wave have begun to be non-linearly steepened. This is particularly clear for $a_{0}=1.0, 1.5$ (\Fref{aScan}(b,c)), where the plasma wave consists of sharp density spikes either side of shallow long density troughs. As a result, the electric field becomes sawtoothed in nature, with a long region of linearly increasing field followed by a sharp drop due to the density spikes.
In the non-linear case too, the fluid approximation breaks down for large amplitudes, as particle sheets cross from neighbouring wave buckets, leading to wavebreaking. Due to the non-linear steepening, the density spikes have much higher density than in the linear case, and the value at which wavebreaking occurs (for a cold relativistic plasma) is given by \cite{akhiezer} \[E_\mathrm{WB} = \sqrt{2}(\gamma_\mathrm{ph}-1)^{1/2} E_{0}.\]
The non-linearities in density also causes lengthening of the plasma wave. This can be seen in \Fref{aScan2}(a), where adding relativistic effects results in an increasing lengthening of the laser pulse length at which resonance occurs. This is because the plasma wave wavelength increases as relativistic effects give the plasma electrons greater inertia. Figure \ref{aScan2}(b) shows the increase of the maximum electric field with increasing laser strength $a_{0}$. At low intensity, the field increases proportional to $a_{0}{^{2}}$, whereas at high laser strength the dependence becomes closer to $a_{0}$. This dependence follows the ponderomotive potential of the laser pulse. Because of the non-linear steepening, the electric field can easily exceed the cold wavebreaking limit $E_{0}$ at high $a_{0}$. At resonance, the wake amplitude is given by

$$E_\mathrm{max} \approx \frac{{a_{0}}^{2}}{(1+{a_{0}}^{2})^{1/2}}E_{0}.$$

\section{Electromagnetic waves}

\subsection{The wave equation in vacuum}

The wave equation in vacuum is
\begin{equation*}
\mathbf{\nabla^{2} E} - \frac{1}{c^{2}} \frac{\partial^{2} \textbf{E}}{\partial t^{2}} =0.
\end{equation*}
The \emph{paraxial ray approximation} assumes that the phase of the electromagnetic wave varies primarily in $z$, the direction of propagation, i.e.~$\textbf{E} = E(x,y,z) \exp(\mathrm{i}(kz-\omega t)) \hat{\textbf{x}}$, which is true for gentle focussing as in wakefield accelerators.
So,
\[ \nabla^{2} \textbf{E} = \left(\frac{\partial^{2}}{\partial x^{2}} + \frac{\partial^{2}}{\partial y^{2}} \right) \textbf{E}  + \frac{\partial^{2} E}{\partial z^{2}} \mathrm{e}^{\mathrm{i}(kz-\omega t)}  \hat{\textbf{x}} + \mathrm{i}k \mathrm{e}^{\mathrm{i}(kz-\omega t)}\frac{\partial E}{\partial z} \hat{\textbf{x}} - k^{2} \textbf{E}.
\]
The last term, $k^{2}\textbf{E}$, cancels with $\displaystyle \frac{1}{c^{2}} \frac{\partial^{2} \textbf{E}}{\partial t^{2}} = \frac{\omega^{2}}{c^{2}}\textbf{E}$ since $\omega \approx ck$, and we can assume that the variation in $z$ is slow, so that $\displaystyle \frac{\partial^{2} E}{\partial z^{2}} \rightarrow 0$, leading to
\[ \left(\frac{\partial^{2}}{\partial x^{2}} + \frac{\partial^{2}}{\partial y^{2}} \right) E - 2\mathrm{i}k \frac{\partial E}{\partial z} = 0. \tag{cartesian paraxial ray wave equation} \]
In two dimensions, e.g., taking $\displaystyle \frac{\partial^{2} E}{\partial y^{2}} =0 $, the equation forms a time-dependent Schr\"odinger equation.

\subsection{Gaussian optics}

\begin{figure}[h]
\begin{center}
\includegraphics[width=12cm]{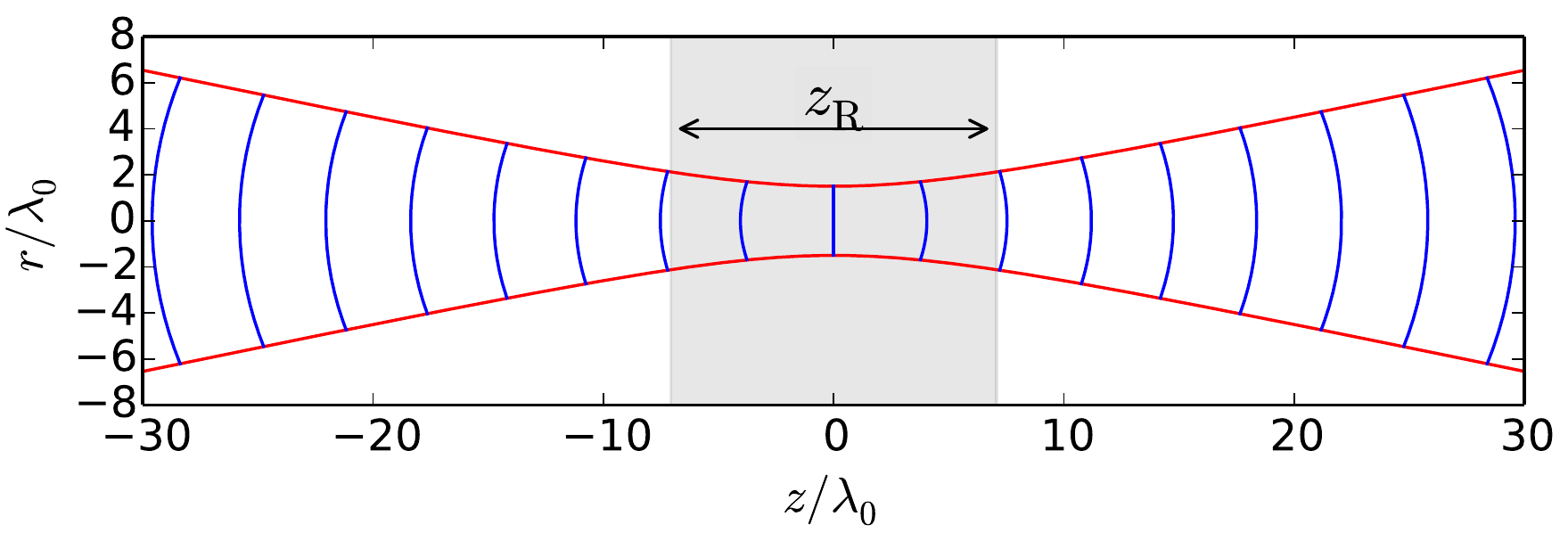}
\caption{Evolution of beam waist (red line) of an ideal Gaussian beam, with curvature of phase fronts (blue lines) and Rayleigh range, $z_\mathrm{R}$, (shaded grey) also shown.}
\label{gaussOpt}
\end{center}
\end{figure}

For a cylindrically symmetric beam, the paraxial ray equation can be rewritten:
\[ \frac{\partial^{2} E}{\partial r^{2}} + \frac{1}{r}\frac{\partial E}{\partial r} - 2\mathrm{i}k \frac{\partial E}{\partial z} = 0. \tag{paraxial ray wave equation} \]
This equation has a solution of the form:
$$E(r,z)= \frac{w_{0}}{w} \exp \left[\frac{-r^{2}}{w^{2}} - \frac{\mathrm{i}\pi r^{2} }{\lambda R} + \mathrm{i}\phi_{0} \right],$$
where
\[ w =  w_{0} \sqrt{1+ \left(\frac{z}{z_\mathrm{R}}\right)^{2}}, \tag{beam waist} \]
\[ R=\frac{1}{z} \left( z^{2} + {z_\mathrm{R}}^{2} \right), \tag{radius of curvature}\]
\[ \tan \phi_{0} = \frac{\lambda z}{\pi {w_{0}}^{2}}, \tag{Gouy phase} \]
where $\displaystyle z_\mathrm{R} = \frac{\pi {w_{0}}^{2}}{\lambda}$ is the Rayleigh length, and
$w_{0}$ is the beam waist, and its variation in vacuum can be seen in the \Fref{gaussOpt}. The beam has a minimum at $z=0$ and like an unconstrained wave packet, will disperse with increasing propagation. $R$ is the curvature of the incoming phase fronts and $\phi_{0}$ is known as the Gouy phase, which flips through zero as a beam passes through focus.

The intensity is then given by $\displaystyle {I}(r,z)/ I_0 = \frac{w_{0}^2}{w^2} \exp \left[\frac{-2r^{2}}{w^{2}} \right]$, so for $z = \pm z_\mathrm{R}$, the intensity falls by half. As can be seen in \Fref{gaussOpt}, this is also the distance over which the phase fronts are approximately flat. For $z\gg z_\mathrm{R}$, the beam waist expands almost linearly with an angle given by $$\tan \theta = \lim_{z\gg z_\mathrm{R}} \left[\frac{w}{z} \right] = \frac{w_{0}}{z_\mathrm{c}} = \frac{\lambda}{\pi w_{0}}.$$
Or $\displaystyle \theta \sim \frac{1}{\pi F}$, where the $F$-number, $\displaystyle F = \frac{w_{0}}{\lambda} = \frac{f}{d}$, which is the ratio of focal length, $f$, to the beam diameter at the final mirror, $d$.

For wakefield generation, it would seem advantageous to have as high an intensity as possible, i.e.~focusing to small focal spot size. This however implies a short Rayleigh range and so a short acceleration distance. To try to maintain a near constant intensity over the interaction distance ($z_\mathrm{R}>L_\mathrm{deph}$) would require gentle focussing. This has an implication though for the real size of a plasma accelerator, since the beam diameter is typically constrained by damage thresholds on the final turning mirror ($I_\mathrm{dam}\sim 10^{12}$ W cm$^{-2}$), which then has to be further away from final focus for a longer focal length interaction. Luckily, plasma effects can help guide an intense laser beam for distances much longer than the Rayleigh range, alleviating this problem.

\subsection{Propagation in plasma}

The wave equation including the effect of plasma can be written
\begin{equation*}
\mathbf{\nabla^{2} E} - \frac{\eta^{2}}{c^{2}} \frac{\partial^{2} \textbf{E}}{\partial t^{2}} =0,
\end{equation*}
where $\eta$, the plasma refractive index, is given by
$$\eta_\mathrm{R} \simeq 1 - \frac{\omega_\mathrm{p}{^{2}}}{2\omega^{2}} \frac{n(r)}{n_{0}\gamma_{\perp}},
$$
for the case of sufficiently underdense plasmas, i.e. ${\omega_\mathrm{p}{^{2}}}/{\omega^{2}} \ll 1$. Note that both the density profile $n(r)$ and the relativistic factor $\gamma_{\perp}(r)$ can now vary radially, in the latter case due to the dependence of $\gamma_{\perp}$ on intensity, i.e. $\gamma_{\perp} = \sqrt{1 + (a_0^2/2)}$ for linear polarization.

Either an intensity decrease away from the axis of propagation ($\partial a^{2}/\partial r^{2}) < 0$ or a density increase ($\partial n_\mathrm{e}^{2}/\partial r^{2} > 0$) can lead to a higher refractive index on-axis and so, like an optical fibre, causes flow of energy towards the axis or self-focusing. For small variations, the refractive index can be written
$$\eta_\mathrm{R} \simeq 1 - \frac{\omega_\mathrm{p}{^{2}}}{2\omega^{2}} \left(1 + \frac{\delta n}{n_{0}} - \frac{a^{2}}{2}\right),$$
where $\delta n$ is the density depression, which can be due to a number of effects such as a preformed channel, or ponderomotive expulsion of plasma electrons.

\begin{figure}[h]
\begin{center}
(a) \includegraphics[width=4.6cm]{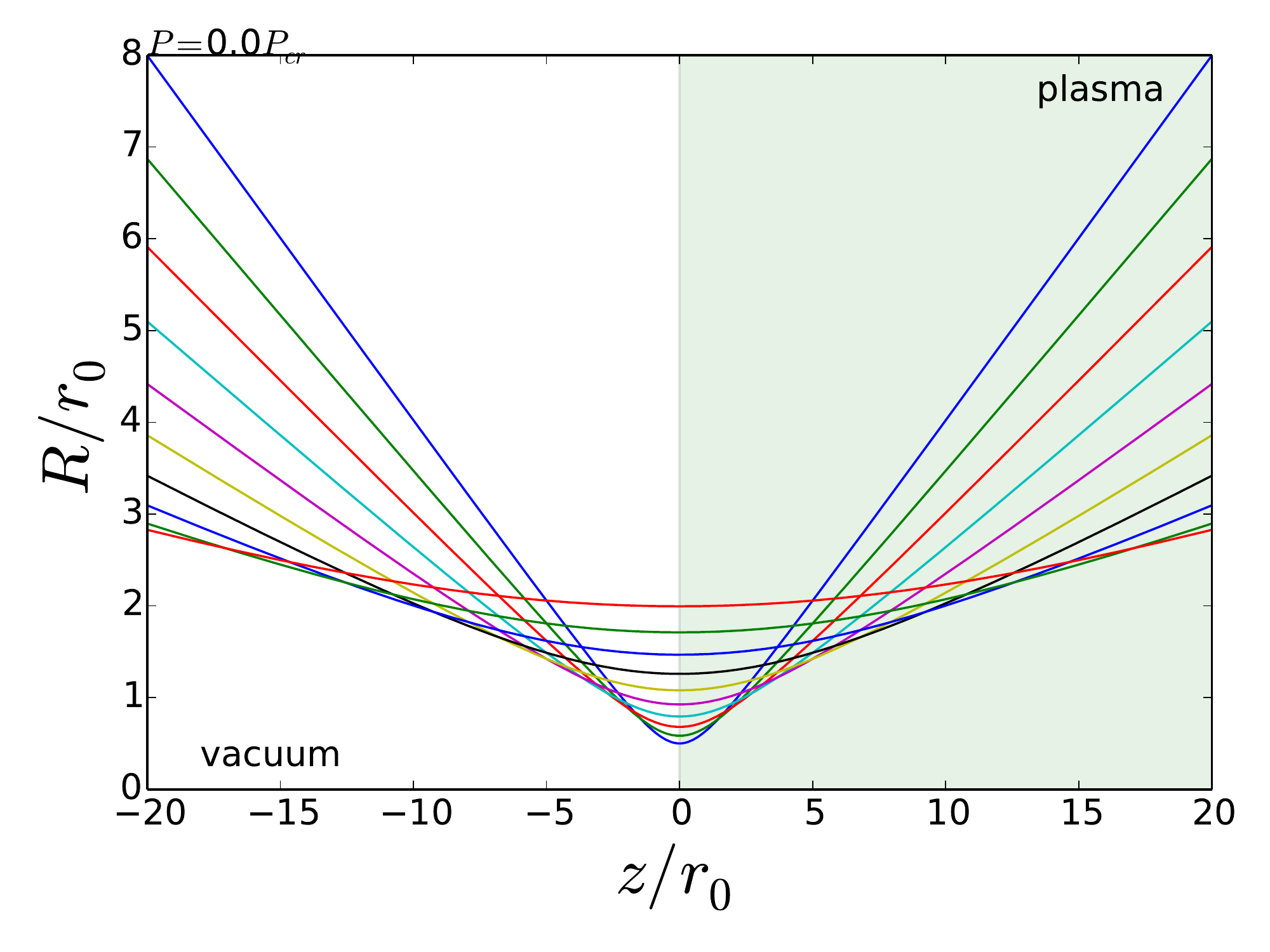}
(b) \includegraphics[width=4.6cm]{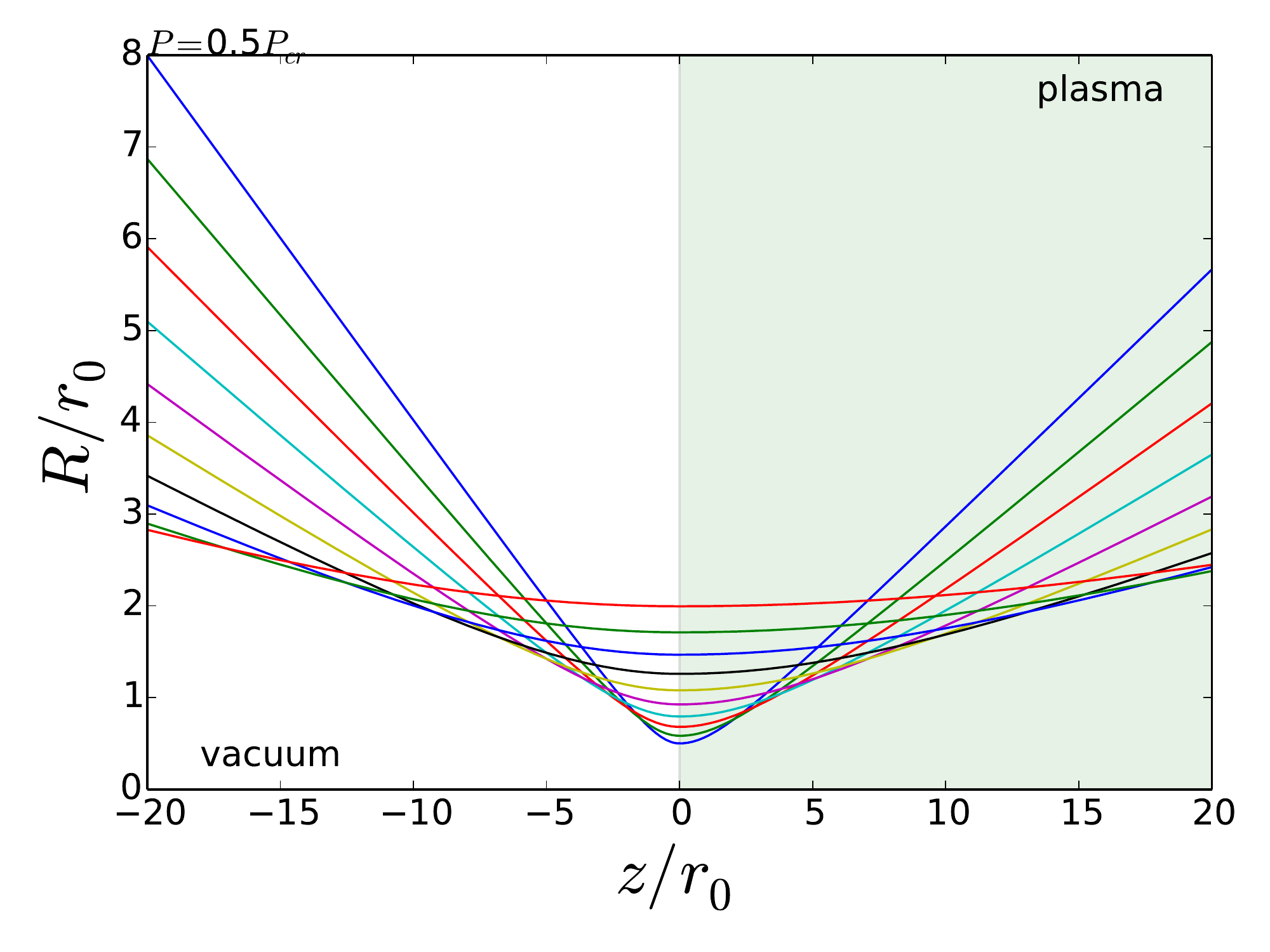}
(c) \includegraphics[width=4.6cm]{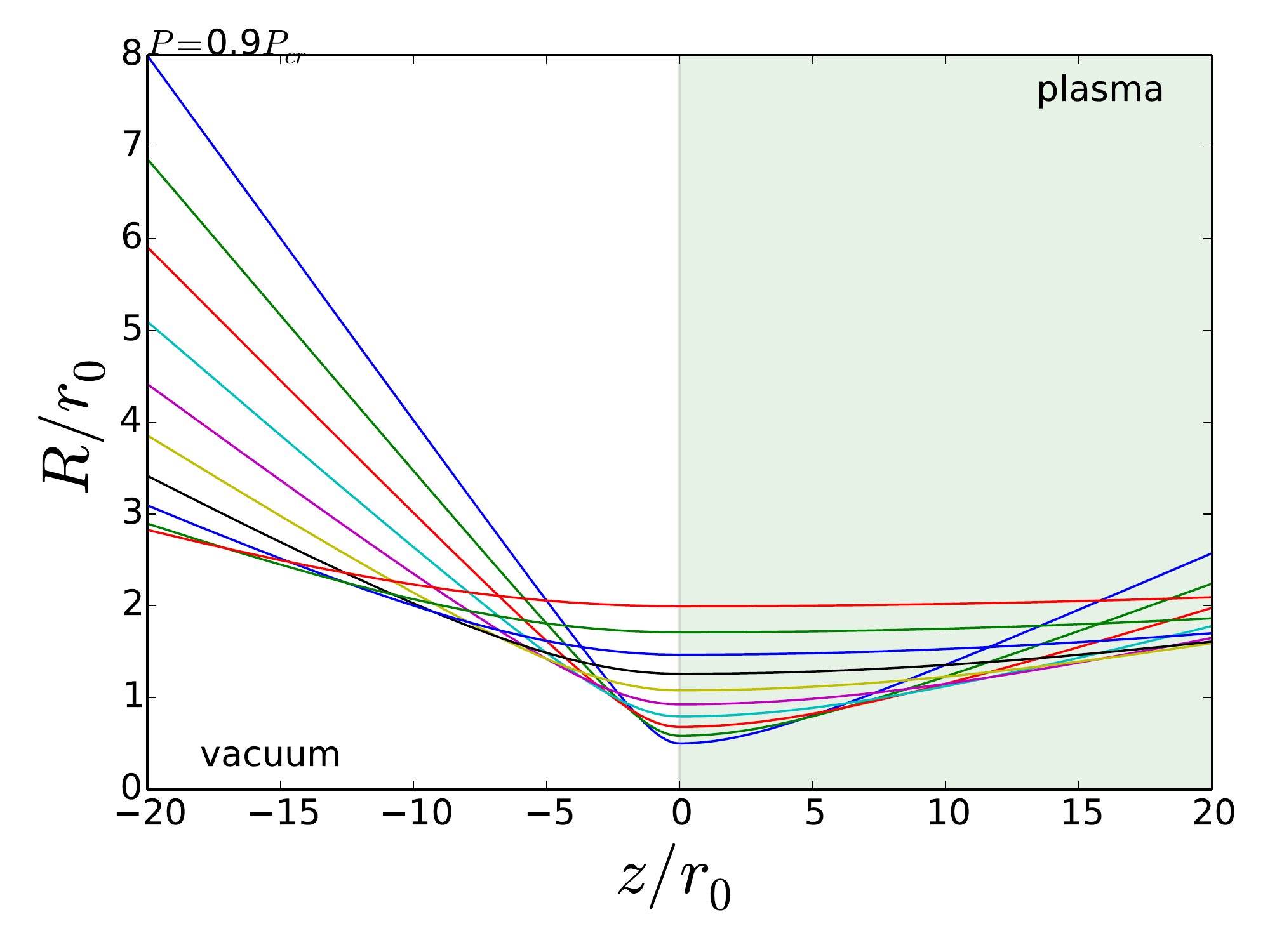}
(d) \includegraphics[width=4.6cm]{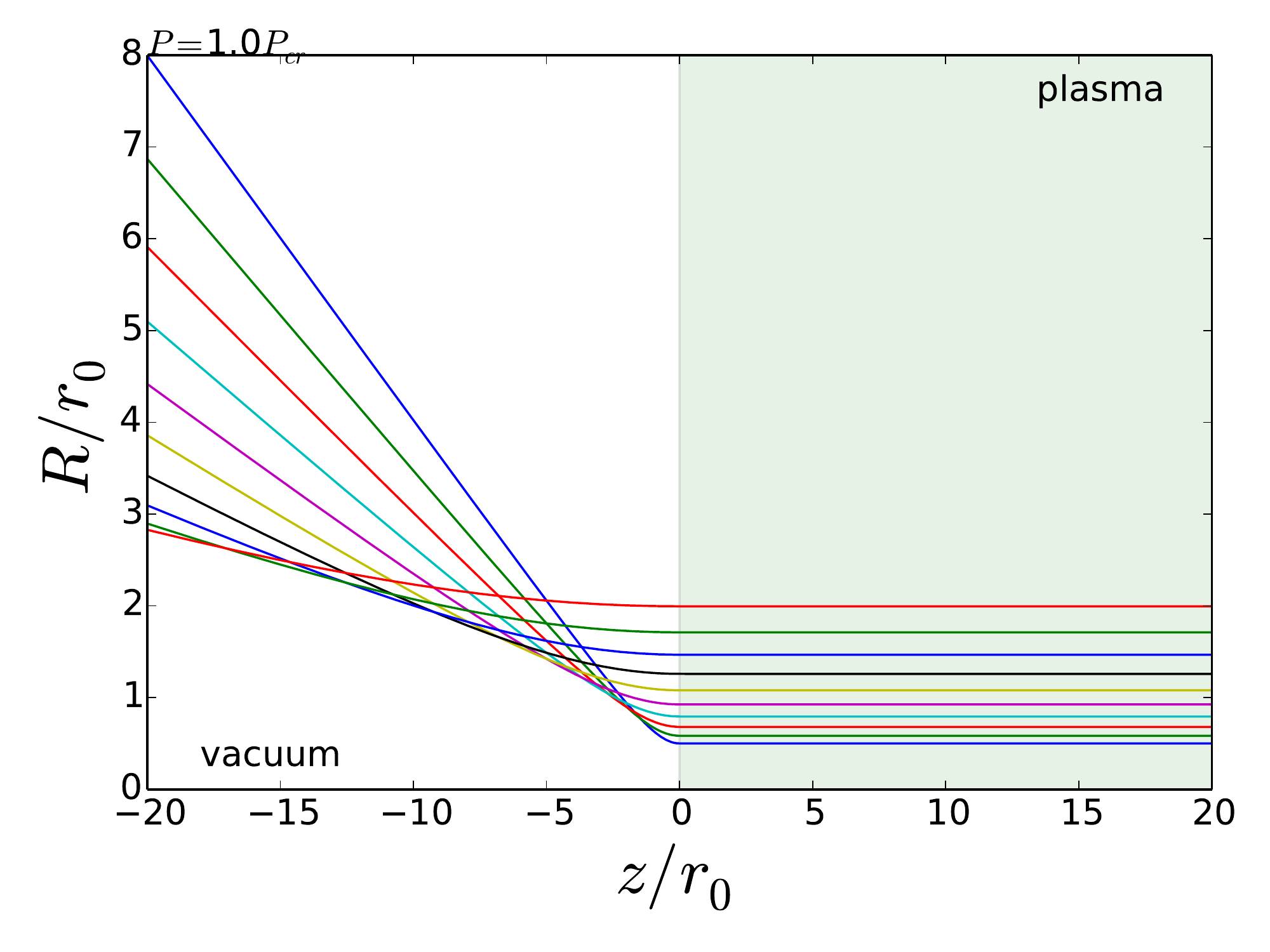}
(e) \includegraphics[width=4.6cm]{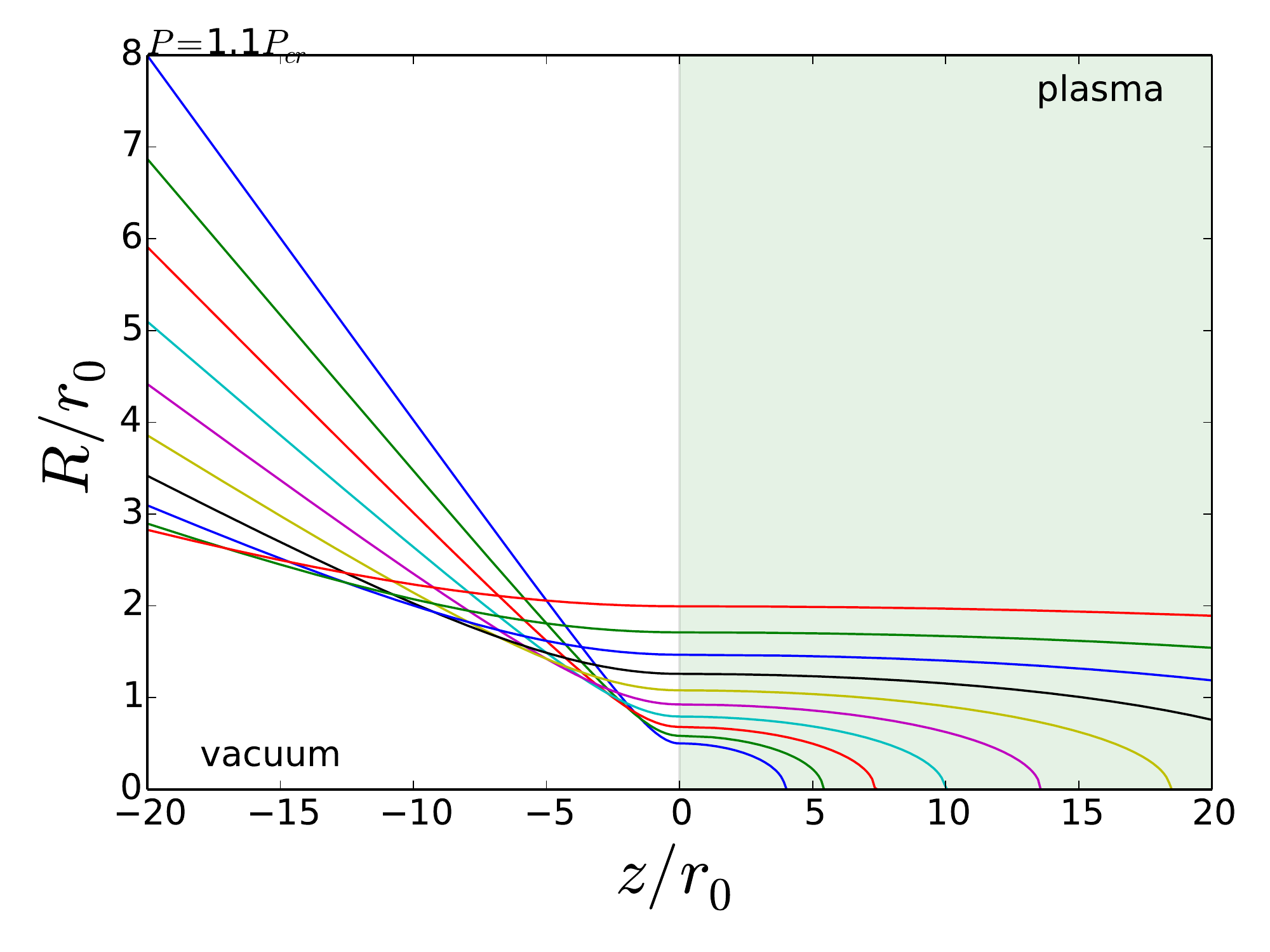}
\caption{Behaviour of laser beam waist focussed on boundary of sharp density plasma transition for  $P/P_\mathrm{cr} =$ (a) 0, (b) 0.5, (c) 0.9, (d) 1.0 and (e) 1.1.}
\label{selfFoc}
\end{center}
\end{figure}

A Gaussian laser pulse with transverse variation $a(r) = a_{0} \exp(-r^{2}/R^{2})$, where $R$ is now the beam waist, can be used as a trial solution to the paraxial ray equation. Keeping only dominant terms, the evolution of the beam waist can be approximated by \cite{esarey}
$$
\frac{\mathrm{d}^{2} R}{\mathrm{d}z^{2}} = \frac{1}{z_\mathrm{R}{^{2}}R^{3}}\left(1- \frac{P}{P_\mathrm{c}} \right).
$$
The spot size has been normalized to $r_{0}$, the minimum vacuum focal spot size of the laser pulse and the power unit $P_\mathrm{c} \simeq 17({\omega{^{2}}}/{{\omega_\mathrm{p}}^{2}})$\, GW, which is dependent on density. Figure~\ref{selfFoc} demonstrates the effect of $P/P_\mathrm{cr}$ on the evolution of the beam waist. Increasing $P/P_\mathrm{cr}$ leads to a greater influence of the second term, which is due to intensity effects, over the first term which causes diffraction. At $P = P_\mathrm{cr}$, a balance is formed between diffraction and self-focusing and the laser pulse can be guided. Hence this is a \emph{critical} power for self-focusing. For $P>P_\mathrm{cr}$, the model predicts catastrophic self-focusing. In reality however, including higher order diffraction terms prevents this as strong focussing invariably leads to generation of higher modes and thus filamentation \cite{thomas}.

\subsection{Guiding}

For wakefield driving short pulses, the leading edge of the pulse interacts with the wrong curvature of density due to the plasma wave (increasing on-axis) and this can cause increased diffraction of the laser pulse. This is true even if the body of the laser pulse is being self-focused. To overcome this, guiding channels are used, where the guiding of the pulse is now determined by
$$\frac{\mathrm{d}^{2} R}{\mathrm{d}z^{2}} = \frac{1}{z_\mathrm{R}{^{2}}R^{3}}\left(1- \frac{\delta n}{\delta n_\mathrm{c}} R^{4} \right).$$
A Gaussian beam with $R=1$, i.e. $r = r_{0}$, can be guided, provided there is a density depression with $\delta n = \delta n_\mathrm{c}$, where
\begin{equation}
\delta n_\mathrm{c} = \frac{1}{\pi r_\mathrm{e} r_{0}{^{2}}},
\label{eqn:matched}
\end{equation}
and $r_\mathrm{e} = e^{2}/m_\mathrm{e}{^{2}}c^{2}$ is the `classical radius of the electron'. Such channels can be formed, for example, by thermal equilibration of a discharge with its cold walls, or by ponderomotive or thermal expulsion of a plasma by a secondary laser pulse \cite{esarey2}. Channel guiding allows laser energy to be used more efficiently in a laser wakefield and also allows potentially simpler operation in the linear regime.

\begin{figure}[h]
\begin{center}
(a) \includegraphics[width=4.6cm]{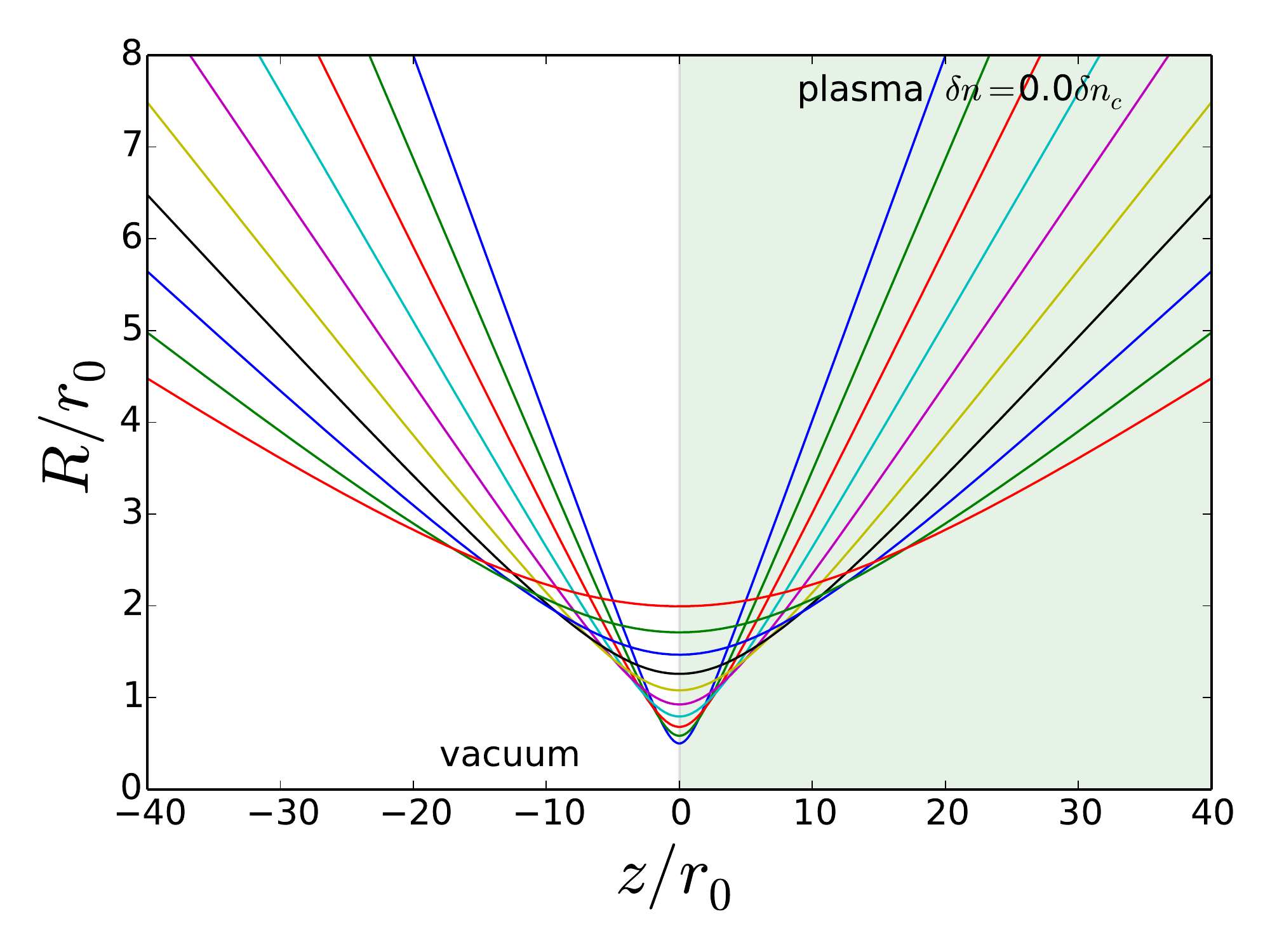}
(b) \includegraphics[width=4.6cm]{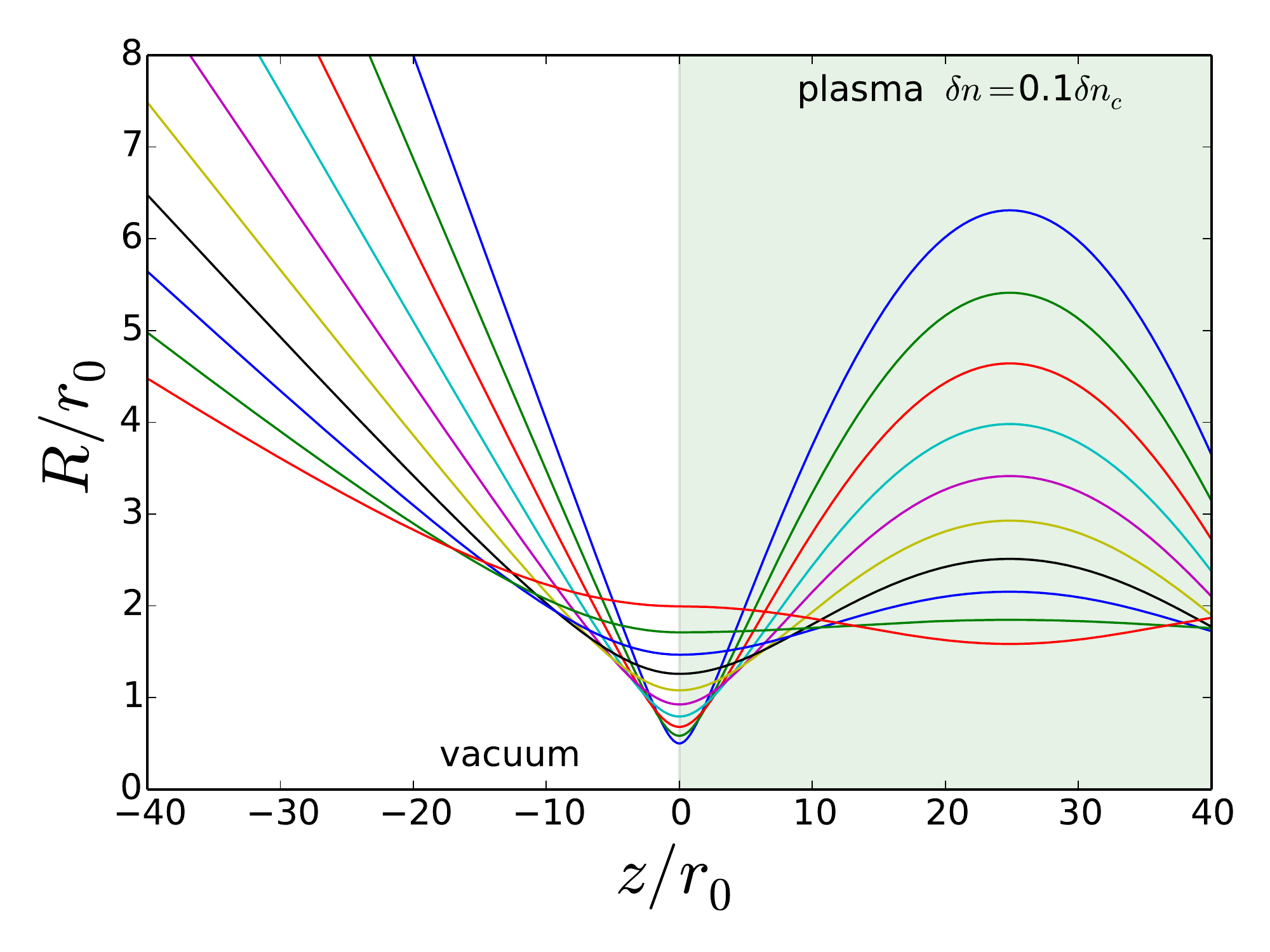}
(c) \includegraphics[width=4.6cm]{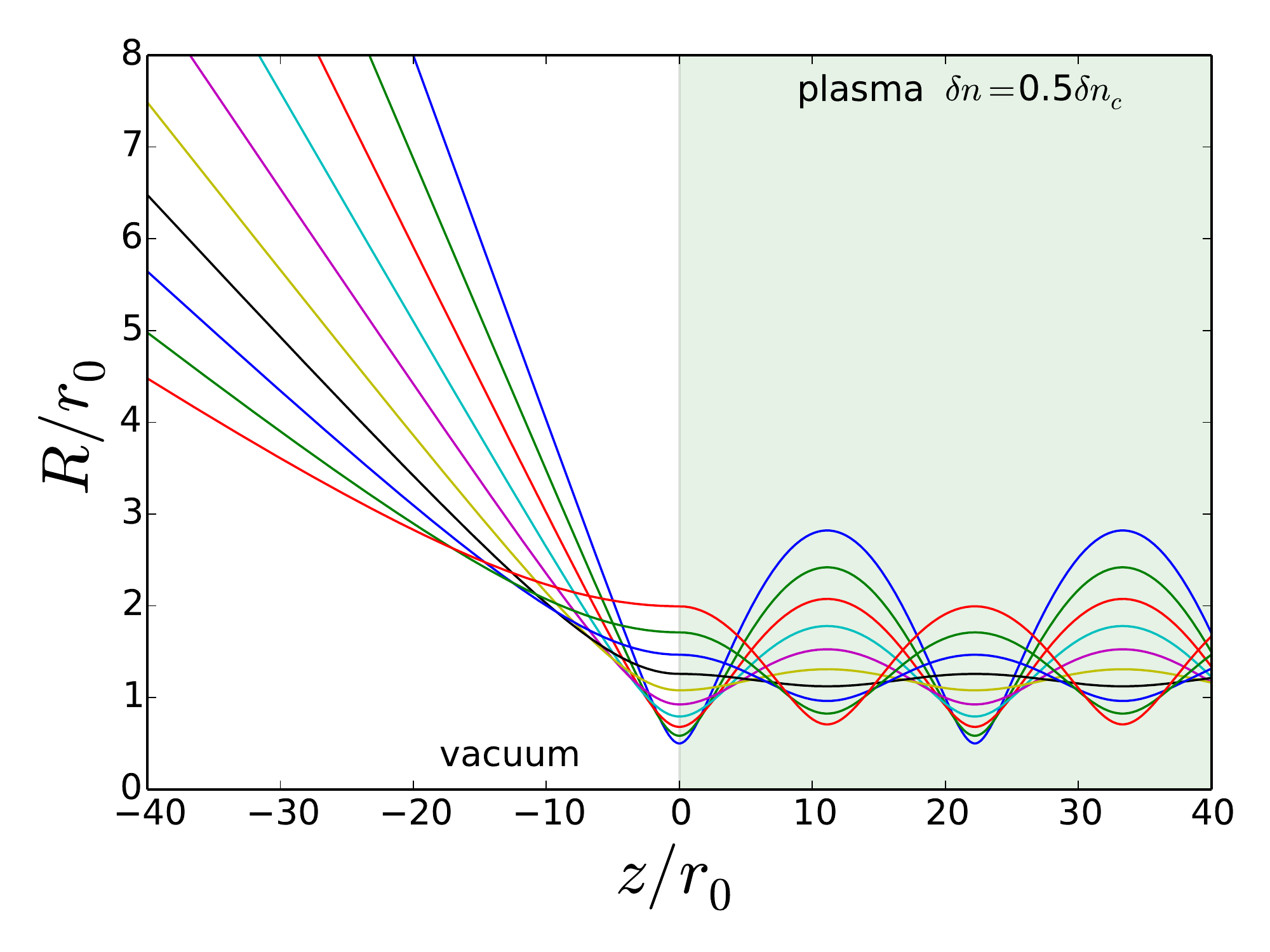}
(d) \includegraphics[width=4.6cm]{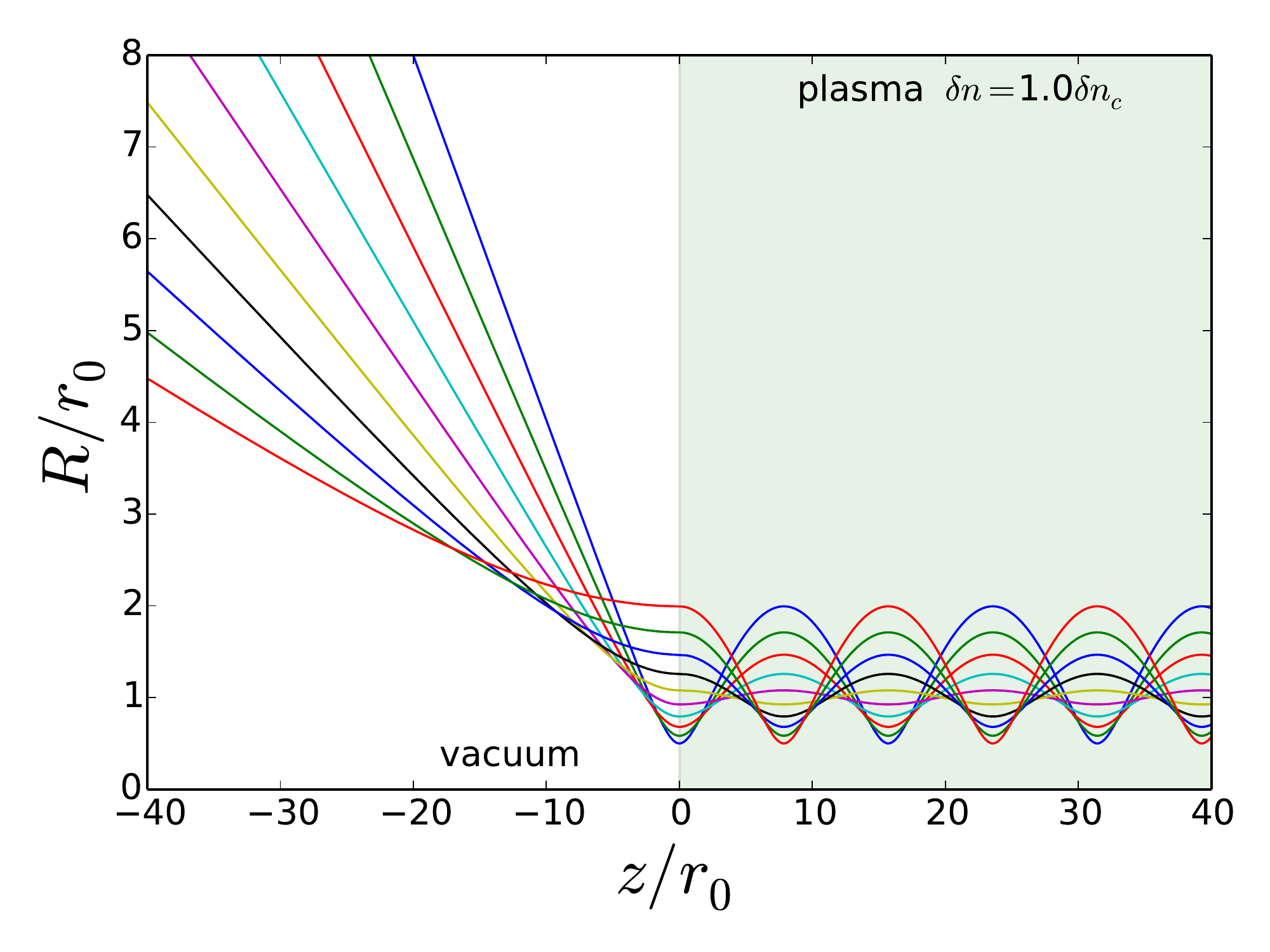}
(e) \includegraphics[width=4.6cm]{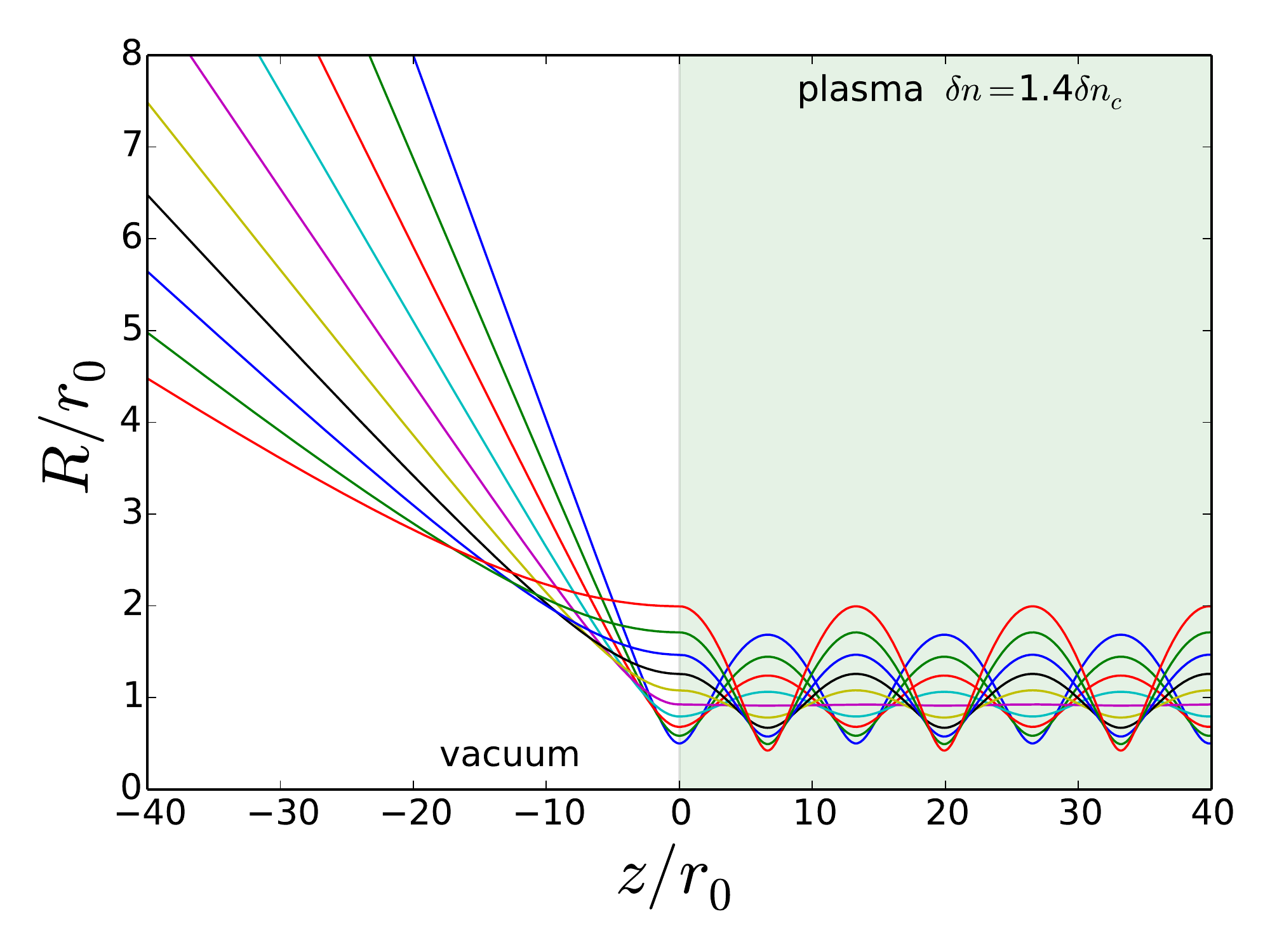}
\caption{Behaviour of the waist of the leading edge of a laser pulse focussed on boundary of sharp density transition on a plasma with radial guiding channel with  $\delta n/\delta n_\mathrm{cr} =$ (a) 0, (b) 0.1, (c) 0.5, (d) 1.0 and (e) 1.4.}
\label{channelFoc}
\end{center}
\end{figure}

There is a critical beam or `matched' spot size at which the beam propagates with minimum channel waist oscillations (see \Fref{channelFoc}), as can be found by inverting \Eref{eqn:matched}. The goal for laser wakefield operation is to match the Gaussian focussed laser spot to the matched spot of the accelerator, which determines the required focussing parameter. However, as can be seen in the figure, not focussing at the matched spot size will lead to channel oscillations which can lead to a loss of guided energy, as the channels are invariably `leaky'.

\subsection{Propagation effects}

\begin{figure}[h]
\hspace{2.2cm} (a) \hspace{5.3cm} (b) \vspace{-3mm}
\begin{center}
\includegraphics[width=6cm]{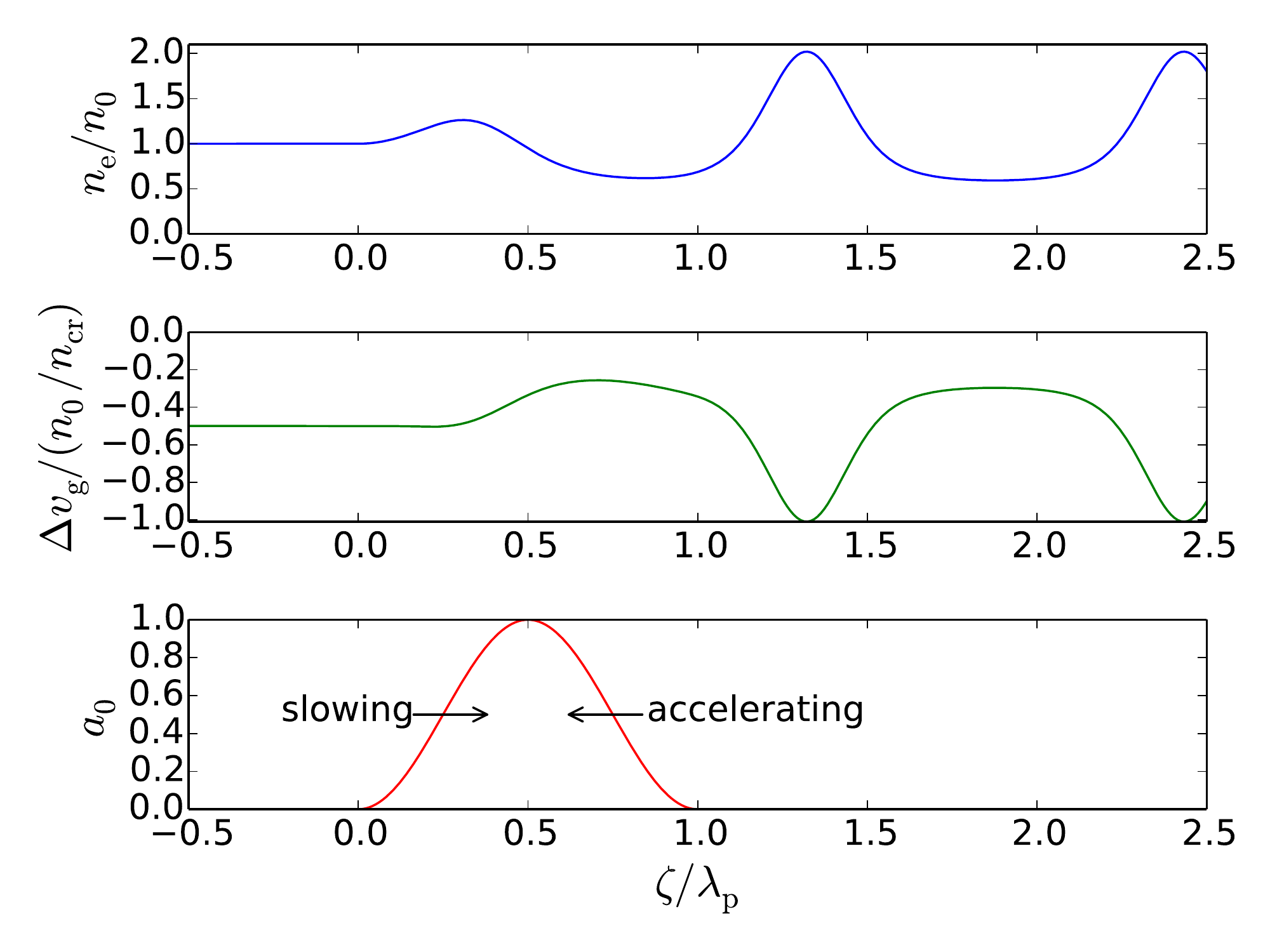}
\includegraphics[width=6cm]{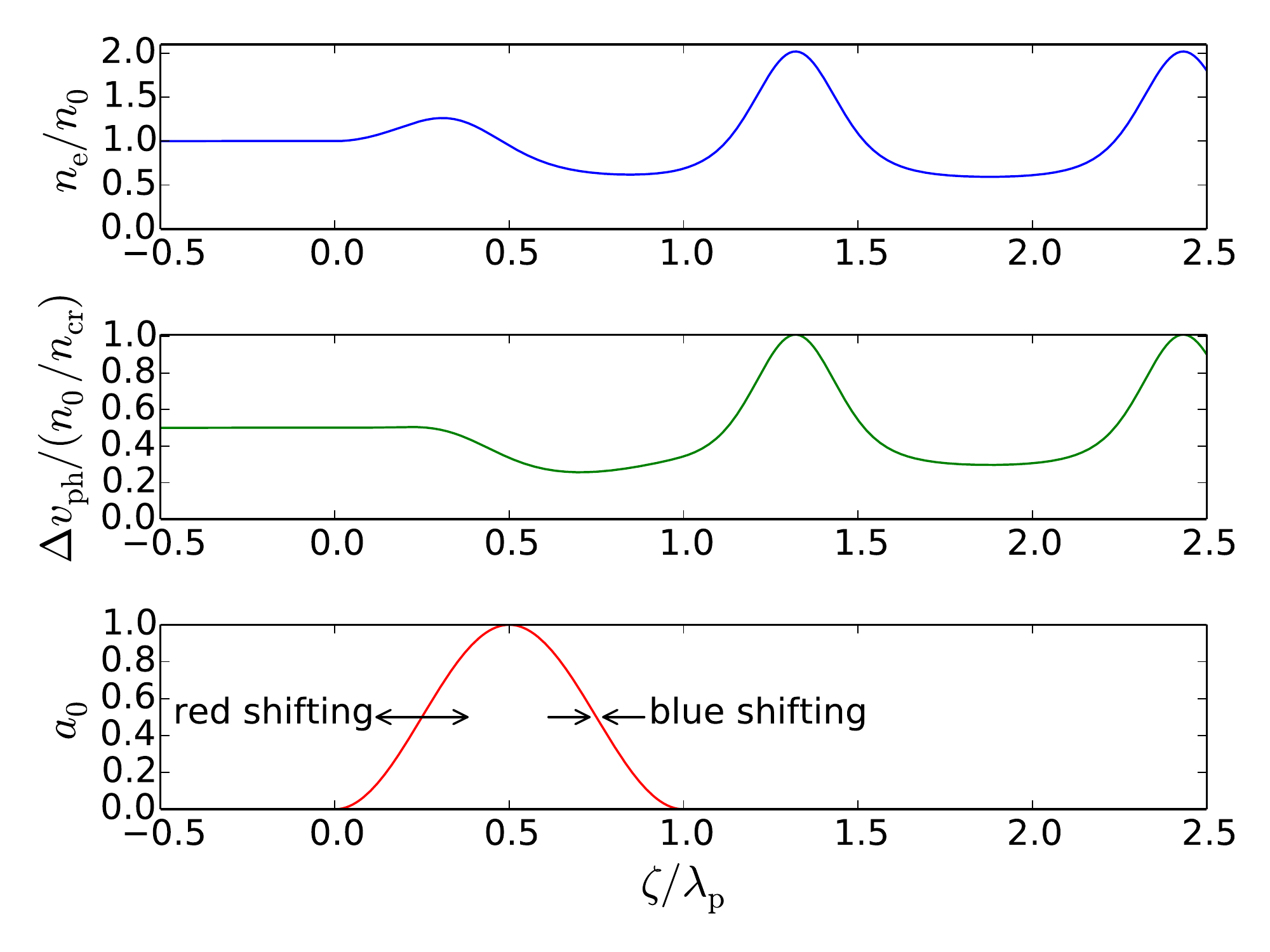}
\caption{Compression and photon acceleration of a wakefield driving laser pulse for $a_{0} =1$, $L=\lambda_\mathrm{p}$. (a) $n_\mathrm{e}$ (top), \emph{group} velocity (middle) and pulse shape (bottom) showing regions of speeding and slowing of the pulse. (b) $n_\mathrm{e}$ (top), \emph{phase} velocity (middle) and pulse shape (bottom) showing regions of stretching and compressing of phase.}
\label{compress}
\end{center}
\end{figure}

A laser pulse of course does not stay unchanged as it drives a wakefield, it must lose energy. However this does not necessarily lead to a reduction in intensity, since the pulse can compress as it propagates. In a linear wake, the rate of compression is given by the difference in the group velocity from the first maximum to minimum of the plasma wave.
This can be seen in \Fref{compress} which shows that due to the density rise, the front of the laser pulse has a slower group velocity than the rear of the pulse, causing the back to `catch-up' and the pulse to compress. For the case of a pulse with $L \sim \lambda_\mathrm{p}$, the variation of pulse length as a function of propagation distance $\ell$ is \cite{schreiber} \[ \tau = \tau_0 - \frac{n_{e0}l}{2n_\mathrm{cr}c}. \]
The pulse compression has the beneficial effect that the power of the driving pulse can stay relatively constant, even though the energy within the laser pulse is being progressively reduced.
The laser pulse also redshifts from the front as those photons that drive the wakefield lose energy. This almost complete loss of energy of the driving photons leads to an etching of the laser pulse. By contrast the trailing photons in the wake are `squeezed' as they travel in the density depression and so become blue shifted. This process is often called photon acceleration and the amount of blue shifting (energy gain) of the photon is given by \cite{wilks}
$$
\delta \omega = \omega_0\left(1-z\frac{\mathrm{d}\beta_\mathrm{p}}{\mathrm{d}\zeta}\right) \simeq \omega_0\left(1-z\frac{\mathrm{d}}{\mathrm{d}\zeta}\left(\frac{\delta n}{n_0} \right)\right).
$$
Photon acceleration can be a useful diagnostic of wakefields in cases where there are no charged particles available to accelerate \cite{murphy}.

\section{Conclusion}
We have discussed a number of effects which are important for laser wakefield acceleration. We have shown how non-linear effects increase the plasma wave wavelength and also allow larger wakefield amplitudes at high intensities ($a_{0}>1$). This is due to the characteristic peak and trough nature of the non-linear wakes in this regime.

We have also discussed simple models for the guiding of laser pulses which allow the length of a laser wakefield well beyond the Rayleigh range of the laser pulse. This is important as it allows the full energy gain to be extracted from the laser wakefield. We also discussed how preformed channels can enhance the guiding effect for a laser pulse focussed to the matched spot size of the channel.

\section*{Bibliography}

P. Gibbon, \emph{Short Pulse Laser Interactions with Matter}
(World Scientific Publishing Company, New York, 2004).

\end{document}